\documentclass[arguments]{aastex631}
\shorttitle{Diversity of dust properties in external galaxies}
\shortauthors{Nagao et al.}
\graphicspath{{./}{figures/}}

\usepackage{color}

\begin{document}

\title{Diversity of dust properties in external galaxies confirmed by polarization signals from Type II supernovae}

\correspondingauthor{Takashi Nagao}
\email{takashi.nagao@utu.fi}

\author[0000-0002-3933-7861]{Takashi Nagao}
\affiliation{Department of Physics and Astronomy, University of Turku, FI-20014 Turku, Finland}

\author[0000-0002-0537-3573]{Ferdinando Patat}
\affiliation{European Southern Observatory, Karl-Schwarzschild-Str.\ 2, 85748 Garching b. M\"{u}nchen, Germany}

\author[0000-0003-2611-7269]{Keiichi Maeda}
\affiliation{Department of astronomy, Kyoto University, Kitashirakawa-Oiwake-cho, Sakyo-ku, Kyoto 606-8502, Japan}

\author[0000-0003-1637-9679]{Dietrich Baade}
\affiliation{European Southern Observatory, Karl-Schwarzschild-Str. 2, 85748 Garching b. M\"{u}nchen, Germany}

\author[0000-0002-8255-5127]{Seppo Mattila}
\affiliation{Department of Physics and Astronomy, University of Turku, FI-20014 Turku, Finland}
\affiliation{School of Sciences, European University Cyprus, Diogenes street, Engomi, 1516 Nicosia, Cyprus}

\author[0000-0002-4265-1958]{Stefan Taubenberger}
\affiliation{Max-Planck-Institut f\"{u}r Astrophysik, Karl-Schwarzschild-Str.\ 1, 85748 Garching b. M\"{u}nchen, Germany}

\author[0000-0001-5455-3653]{Rubina Kotak}
\affiliation{Department of Physics and Astronomy, University of Turku, FI-20014 Turku, Finland}

\author[0000-0001-7101-9831]{Aleksandar Cikota}
\affiliation{European Organisation for Astronomical Research in the Southern Hemisphere (ESO), Alonso de Cordova 3107, Vitacura, Casilla 19001, Santiago de Chile, Chile}

\author[0000-0002-1132-1366]{Hanindyo Kuncarayakti}
\affiliation{Department of Physics and Astronomy, University of Turku, FI-20014 Turku, Finland}


\author[0000-0002-8255-5127]{Mattia Bulla}
\affiliation{The Oskar Klein Centre, Department of Astronomy, Stockholm University, AlbaNova, SE-10691 Stockholm, Sweden}

\author[0000-0003-0733-7215]{Justyn Maund}
\affiliation{Department of Physics and Astronomy, University of Sheffield, Hicks Building, Hounsfield Road, Sheffield S3 7RH, UK}




\begin{abstract}

Investigating interstellar (IS) dust properties in external galaxies is important not only to infer the intrinsic properties of astronomical objects but also to understand the star/planet formation in the galaxies. From the non-Milky-Way-like extinction and interstellar polarization (ISP) observed in reddened Type~Ia supernovae (SNe), it has been suggested that their host galaxies contain dust grains whose properties are substantially different from the Milky-Way (MW) dust. It is important to investigate the universality of such non-MW-like dust in the universe. Here we report spectropolarimetry of two highly-extinguished Type~II SNe (SN~2022aau and SN~2022ame). SN~2022aau shows a polarization maximum at a shorter wavelength than MW stars, which is also observed in some Type~Ia SNe. This is clear evidence for the existence of non-MW-like dust in its host galaxy (i.e., NGC~1672). This fact implies that such non-MW-like dust might be more common in some environments than expected, and thus it might affect the picture of the star/planet formation. On the other hand, SN~2022ame shows MW-like ISP, implying the presence of MW-like dust in its host galaxy (i.e., NGC~1255). Our findings confirm that dust properties of galaxies are diverse, either locally or globally. The present work demonstrates that further investigation of IS dust properties in external galaxies using polarimetry of highly-reddened SNe is promising, providing a great opportunity to study the universality of such non-MW-like dust grains in the universe.

\end{abstract}



\section{Introduction} \label{sec:intro}

Clarifying the interstellar (IS) dust properties in external galaxies is crucial not only for inferring the intrinsic properties of astronomical objects but also for understanding the star/planet formation in galaxies beyond the Milky Way (MW), where dust plays important roles in the radiative transfer and chemistry. Dust properties in the MW and the Large/Small Magellanic clouds have been extensively studied through analysis of the extinction in stars and the dust radiation \citep[e.g.,][for a review]{Draine2003}, while in external galaxies they have been investigated in much less detail. 

The existence of non-MW-like dust in external galaxies has been implied by the non-MW-like extinction observed in reddened Type~Ia supernovae (SNe), where smaller total-to-selective extinction ratios \citep[$R_{\rm{V}} \lesssim 2$; \textit{e.g.},][]{Tripp1998, Elias-Rosa2006, Elias-Rosa2008, Krisciunas2006, Kowalski2008, Wang2008, Nobili2008, Folatelli2010, Burns2014, Amanullah2014, Amanullah2015, Cikota2016} compared to the typical values for dust extinction in the MW \citep[$R_V\sim 3.1$; \textit{e.g.},][]{Fitzpatrick2007} have been found. Similarly, for some Type~IIP SNe, \citet[][]{Poznanski2009} reported a steep extinction law ($R_{\rm{V}} < 2$) by analyzing their light curves using an empirical standardization method.

The dust in the line of sight to SNe not only extinguishes, but also polarizes the SN light (interstellar polarization; ISP). This allows us to investigate the properties of the dust in the line of sight from polarimetric observations of SNe, particularly from the empirical relation between extinction and the wavelength of the polarization maximum, $\lambda_{\rm{max}}$, by \citet[][$R_{V} \sim 5.5\lambda_{\rm{max}}$ $\lbrack \mu$m$\rbrack$]{Serkowski1975}. The ISP of redden Type~Ia SNe shows a polarization maximum at shorter wavelengths \citep[$\lambda_{\rm{max}} \lesssim 0.4$~$\mu$m; \textit{e.g.},][]{Patat2015, Zelaya2017} than the typical MW ISP  \citep[$\lambda_{\rm{max}} \sim 0.545$~$\mu$m;][]{Serkowski1975}. A similar property has also been reported for the ISP toward other types of transients, including the Type~Ib/c SN~2005bf \citep[][]{Maund2007} and the optical transient NGC~300~OT2008-1 \citep[][]{Patat2010}. On the other hand, the Type~II SN~1999gi shows a MW-like ISP curve for its host galaxy, characterized by $\lambda_{\rm{max}}= 0.53$~$\mu$m \citep[][]{Leonard2001}. This demonstrates the existence of MW-like dust in its host galaxy. 

It is important to investigate the universality of such non-MW-like dust in the universe. If such non-MW-like dust grains are common at least in some places in external galaxies, it might qualitatively affect the derivation of the intrinsic properties of astronomical objects and the picture of star/planet formation in galaxies. Given that the existence of non-MW-like dust has been inferred mainly for Type~Ia SN host galaxies and similar investigations for core-collapse SNe have been quite limited, it is important to increase the number of ISP measurements toward core-collapse SNe; by studying them, we may probe the properties of the IS dust in different types of galaxies from the MW and Type~Ia SN host galaxies, or in different environments within the same galaxies. 

In this work, we study the ISP of two reddened Type~II SNe, \textit{i.e.}, SNe~2022aau and 2022ame. SN~2022aau was discovered on 20.60 January 2022 UT during the ongoing $D<40$ Mpc (DLT40) one-day cadence SN search \citep[][]{Tartaglia2018} in NGC~1672 \citep[][]{Bostroem2022a}, located at $z=0.004440$ \citep[][]{Allison2014}. About one day later, the object was classified as a Type~II SN \citep[][]{Siebert2022}. A non-detection of SN~2022aau on 19.56 January 2022 UT, which is about one day before the detection, was reported \citep[][]{Bostroem2022a}. SN~2022ame was discovered on 27.51 January 2022 UT in NGC~1255 \citep[][]{Itagaki2022}, located at $z=0.005624$ \citep[][]{Koribalski2004}. About one day later, the object was classified as a Type~II SN \citep[][]{Bostroem2022b}. The non-detection of SN~2022ame on 24.86 January 2022 UT, which is about three days before the detection, was obtained by the Asteroid Terrestrial-impact Last Alert System (ATLAS) \citep[][]{Tonry2018}.
In the following section, we present details of our observations.
In \S~\ref{sec:results}, we discuss the ISP of these SNe.

\section{Observations and data reduction} \label{sec:obs}

\begin{table*}
\caption{Observation log  and the estimated Serkowski parameters.
}
\label{tab:obs_log}
$
  \begin{tabular}{|c||c|c|c|c||c|c|c|} \hline
    SN & Date (UT) & Phase$^{a}$ & Airmass & Exposure time & $P_{\rm{max}}$ (\%) & $\lambda_{\rm{max}}$ ($\AA$) & $K$ \\ \hline\hline
    SN~2022aau & 2022-01-28.11 & +7.51 & 1.3 & 4$\times$300s & $13.72^{+4.23}_{-0.14}$ & $800^{+90}_{-110}$ & $0.5^{+0.1}_{-0.1}$ \\ \cline{2-5}
     & 2022-03-24.10 & +62.50 & 2.3 & 4$\times$300s & & &\\ \hline
    SN~2022ame & 2022-01-30.10 & +2.59 & 1.4 & 4$\times$600s & $1.44^{+0.01}_{-0.06}$ & $5300^{+190}_{-310}$ & $1.7^{+0.4}_{-0.4}$ \\ \cline{2-5}
     & 2022-03-01.02 & +32.51 & 1.5 & 4$\times$450s & & &\\ \hline
  \end{tabular}
  $

     \begin{minipage}{.88\hsize}
        \smallskip
        Notes. ${}^{a}$Days relative to the discovery. The observational data will be available in the ESO Science Archive Facility at \url{http://archive.eso.org}.
    \end{minipage}
\end{table*}

We have conducted spectropolarimetric observations for SNe~2022aau and 2022ame, using the FOcal Reducer/low-dispersion Spectrograph 2 \citep[hearafter FORS2;][]{Appenzeller1998} instrument mounted on the Cassegrain focus of the Very Large Telescope (VLT) UT1 (Antu) unit telescope in Chile. 
The log of the observations is shown in Table~\ref{tab:obs_log}. We used FORS2 as a dual-beam polarimeter. The spectrum produced by a grism is split by a Wollaston prism into two beams with orthogonal direction of polarization: ordinary (o) and extraordinary (e) beams pass through a half-wave retarder plate (HWP). We used the low-resolution G300V grism coupled to a $1.0$ arcsec slit, giving a spectral coverage of $3800-9200$ {\AA}, a dispersion of $\sim 3.2$ {\AA}~ pixel$^{-1}$ and a resolution of $\sim 11.5$ {\AA} (FWHM) at $5580$ {\AA}. We adopted HWP angles of $0^{\circ}$, $22.5^{\circ}$, $45^{\circ}$ and $67.5^{\circ}$, which are measured between the acceptance axis of the ordinary beam of the Wollaston prism (which was aligned to the north-south direction) and the fast axis of the retarder plate. 

The data were reduced by standard methods with IRAF\footnote{IRAF is distributed by the National Optical Astronomy Observatory, which is operated by the Association of Universities for Research in Astronomy (AURA) under a cooperative agreement with the National Science Foundation.} following \citet[][]{Patat2006}. The ordinary and extraordinary beams were extracted by the PyRAF apextract.apall task with a fixed aperture size of $10$ pixels and then separately binned in $100$ {\AA} bins in order to improve the signal-to-noise ratio. The HWP zeropoint angle chromatism was corrected based on the data in the FORS2 user manual \footnote{\url{http://www.eso.org/sci/facilities/paranal/instruments/fors/doc/VLT-MAN-ESO-13100-1543_P07.pdf}}. The wavelength scale for the Stokes parameters was corrected to the rest-frame using the redshift of the galaxies.

\section{Results and discussion} \label{sec:results}

\begin{figure}[ht!]
\plotone{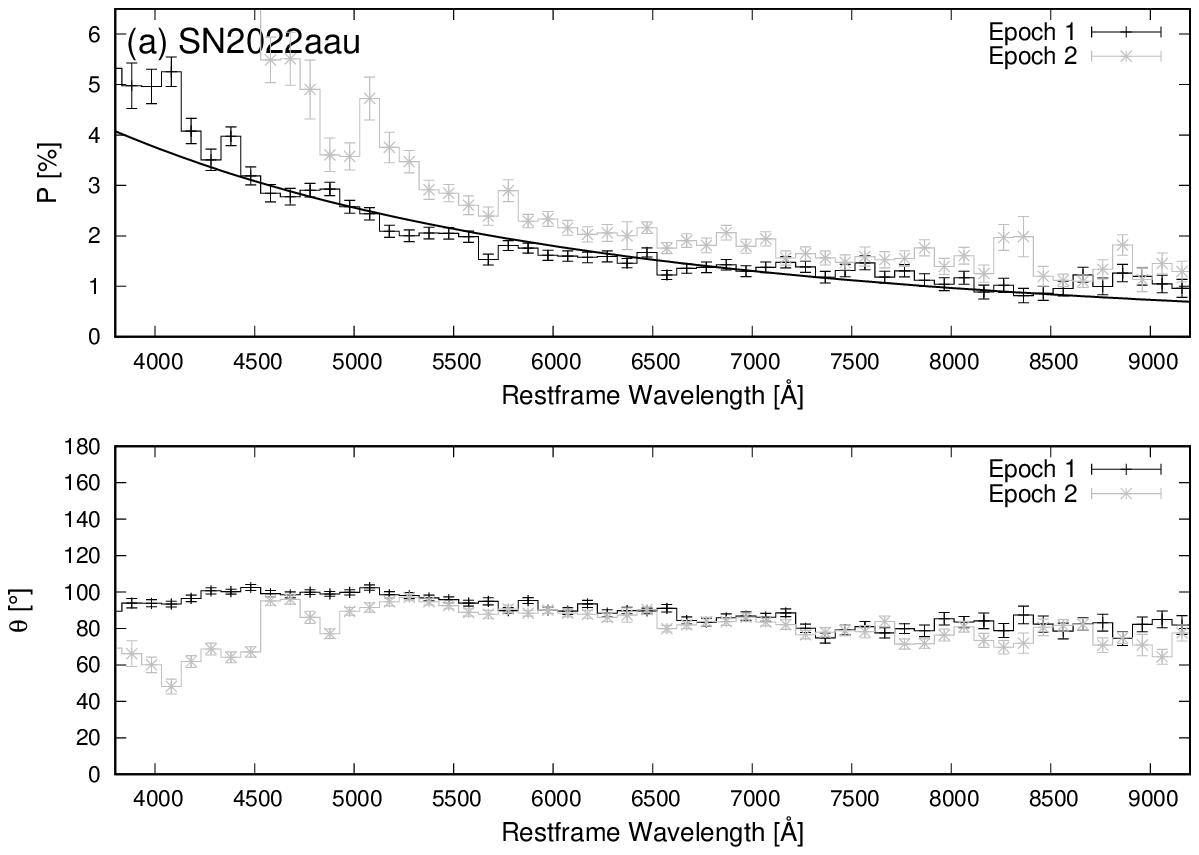}
\plotone{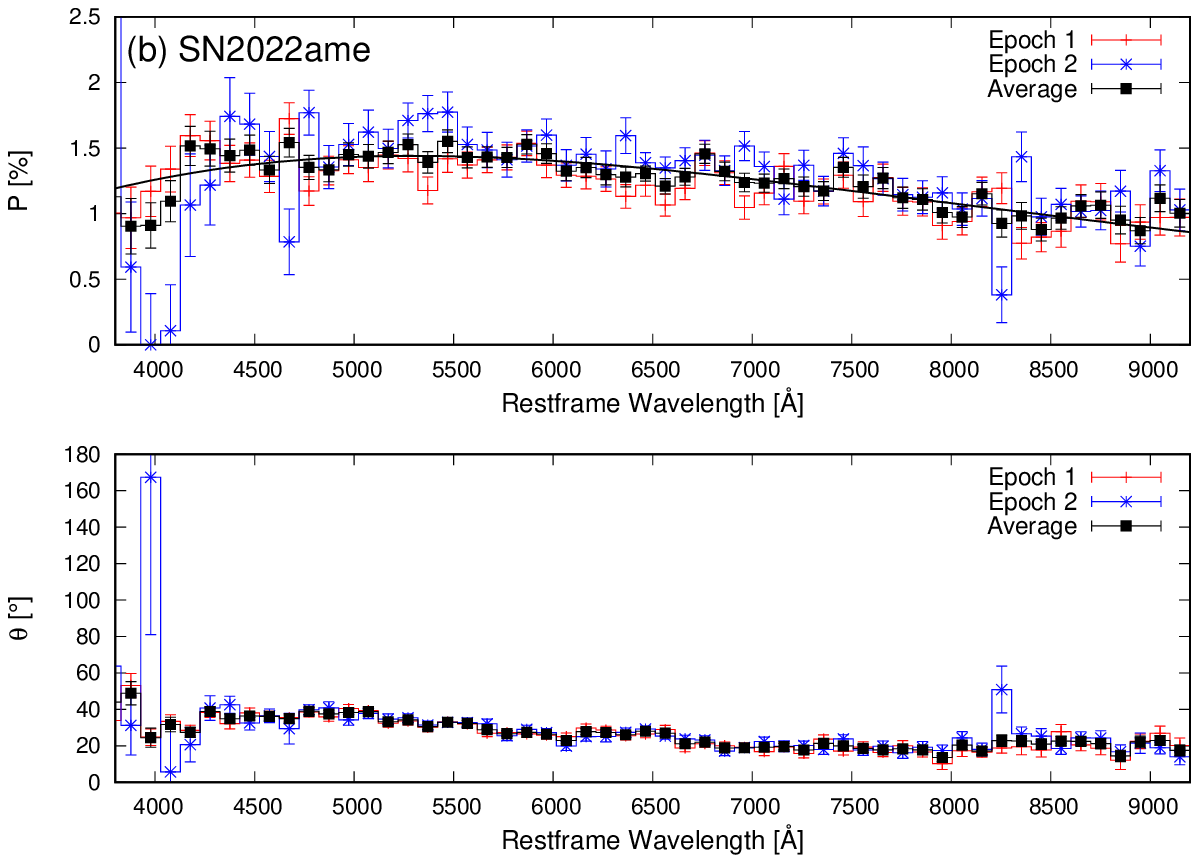}
\caption{Top two panels: Polarization degree $P$ and angle $\theta$ for SN~2022aau at epochs 1 (Phase +8.55 days; black) and 2 (Phase +63.54 days; gray). The black solid line corresponds to the best-fit Serkowski curve. Bottom two panels: Same as upper panels but for SN~2022ame at epochs 1 (Phase +5.24 days; red) and 2 (Phase +35.16 days; blue) as well as the weighted average of the values (black).
}
\label{fig:specpol}
\end{figure}

Figure~\ref{fig:specpol} presents the polarization spectra of SNe~2022aau and 2022ame. SN~2022aau shows a high degree of polarization, i.e., $P\gtrsim 3.0$ \% at $\lambda \sim 4500$ {\AA}, as well as a steep wavelength dependence, i.e., the polarization peaks are at a bluer wavelength than the MW ISP, similar to the ISP of reddened Type~Ia SNe. The spectra at the first and second epochs (Phases +7.51 and +62.50 days) are generally similar, showing a continuum polarization with a single polarization angle of $\sim 90$ degrees. At the same time, the polarization shows a slight increase of the continuum polarization as well as emergence of some line polarization, which corresponds to the line features in the SN spectrum (see Figures~\ref{fig:specpol} and \ref{fig:spec}). Here, we judge that the discrepancy at $\lambda \lesssim 4500$ {\AA} is likely due to the lack of signal, i.e., the incomplete extraction of the spectra (see Fig.~\ref{fig:spec}) and that at the other wavelengths might be due to the intrinsic SN polarization. We will discuss this additional component, probably from the aspherical structure in the SN ejecta, in a forthcoming paper. In the following discussion, we thus use the first-epoch spectrum as a pure ISP component of SN~2022aau. SN~2022ame also shows a high polarisation degree of $P\sim 1.5$ \% at $\lambda \sim 4500$ {\AA}, as well as smooth wavelength dependence
with a peak around $5300$ {\AA}, similar to the MW ISP. There is no noticeable time evolution between the two epochs, and in the following discussion, we therefore use the averaged spectrum from the first and second epochs for SN~2022ame.

Normally, Type~II SNe show low polarization ($\lesssim 0.1$\%) at early photospheric phases \citep[\textit{i.e.}, within a few months after the explosion; \textit{e.g.},][]{WangWheeler2008}, implying that the outermost layers of their progenitors are relatively spherical.  In addition, the polarization that originates from the SN ejecta should have a constant continuum polarization degree through all wavelengths, because the scattering processes in the SN ejecta are dominated by electron scattering, whose opacity is gray \citep[see, e.g.,][]{Nagao2018}. Even in the extreme case of SN~2013ej with a large polarization degree (with no wavelength dependence) just after the explosion, which is interpreted to originate in an aspherical photosphere created by an aspherical circumstellar-material interaction, the continuum polarization was limited to a $\sim0.5$ \% level \citep[][]{Nagao2021}. Such a high intrinsic polarization ($P\gtrsim 1.5$ \%) has not been previously observed in any other Type~II SN at such early phases (within a few months after explosion). Since SNe~2022aau and 2022ame show both a high polarization and a significant wavelength dependence at early phases, their polarization is expected to be imposed externally.

A straightforward interpretation is the ISP, i.e., the polarization due to extinction by aspherical dusts grains aligned in a magnetic field. The Galactic reddening along the lines of sight to SNe~2022aau and 2022ame is $E(B-V)=0.021$ and $0.012$ mag, respectively \citep[][]{Schlafly2011}. With these values, the empirical relation found by \citet[][$P_{\rm{max}}\leq 9 E(B-V)$]{Serkowski1975} suggests that the Galactic ISP for SNe~2022aau and 2022ame should be lower than $\sim 0.2$~\% and $\sim 0.1$~\%, respectively. Therefore, we conclude that the ISP toward SNe~2022aau and 2022ame originates mainly from the dust in their host galaxies. 
Since we do not see the time evolution of the ISP in both SNe, the dust that contributes to the ISP should be located in not a CS scale ($\lesssim 0.1$ pc; where the dust originates from the progenitor systems of these SNe) but a IS scale ($>> 0.1$ pc; where the dust is not directly related with the progenitor systems). The fact that Na~I~D line and the reddening are also constant toward time (see Appendix~\ref{sec:app2}) supports this conclusion.

The other possible external source of polarization with a blue peak is scattering by circumstellar (CS) dust around SNe \citep[e.g.,][]{Patat2005,Nagao2017,Nagao2018}. 
%
However, this scenario is difficult to explain the observed polarization for SNe~2022aau and 2022ame. The polarization degree in this scenario is determined by the relative strength between the SN light and the scattered-echo light. Roughly speaking, to increase the scattered-echo flux towards the SN light, we need to have a larger-solid-angle CSM, which creates more spherical CSM distribution and thus reduces the polarization degree of each scattered photon. As a result, there is an upper limit for the polarization degree that can be created by this scenario toward an assumed input light curve, although, in reality, the echo process is complicated as it depends not only on the time evolution of the central source but also on many other factors, e.g., dust optical properties, the multiple-scattering effects, etc. It has been shown that the polarization level expected in the CSM echo scenario is limited to $\sim 0.1$ \% during the plateau phase of Type IIP SNe, even considering additional factors \citep[][]{Nagao2017,Nagao2018}. In addition, since the scattered echo has a delay time, it cannot contribute to the very early phases as we took spectropolarimetry of SNe~2022aau and 2022ame (Phases +7.51 and +2.59 days, respectively). If we assume that the dust locates just after the dust evaporation radius ($\sim 0.01$ pc for a typical brightness of Type IIP SNe), then the typical delay time is $t_{\rm{delay-time}}=r_{\rm{evp}}/c \sim 10$ days. In the dust scattering scenario, we should also see the time variation of the polarization degree from the early epoch and the latter epoch, which we did not observe for our targets.

SNe~2022aau and 2022ame show substantial reddening in their photometric and spectroscopic properties, and are significantly redder than other Type~II SNe at similar phases (see Fig.~\ref{fig:spec} in Appendix~\ref{sec:app2}). The equivalent widths of the Na~I~D lines formed in the host galaxies of SNe~2022aau and 2022ame also indicate very high extinction: EW$_{\rm{Na I}}$ = 4.8 and 1.4 {\AA} for SNe~2022aau and 2022ame (see Fig.~\ref{fig:spec} in  Appendix~\ref{sec:app2}), respectively, imply $E(B-V) \gtrsim 0.6$ mag based on the empirical relation derived by \citet[][]{Poznanski2012}. This value would be converted into $P_{\rm{max}} \gtrsim 5.4$\% if the above Galactic Serkowski relation is applicable also in these galaxies. The inferred high extinction is in agreement with the high polarization degrees, supporting the conclusion that the ISP within the host galaxies is very likely responsible for the polarization observed toward the two Type~II SNe. 

The ISP angle traces the direction of the magnetic field in the region where the ISP is formed, since the polarization occurs through the differential absorption of the electromagnetic wave by aspherical dust grains aligned with the local magnetic field \citep[\textit{e.g.},][]{Davis1951}. In a spiral galaxy the direction of the magnetic field globally follows the direction of the spiral arms \citep[\textit{e.g.},][]{Beck2015}, even though the magnetic field and thus the ISP might suffer local perturbations, \textit{e.g.}, from supernovae \citep{Ntormousi2018}. The polarization angle in SN~2022aau arguably corresponds to the spiral structure at the location of the SN in its host galaxy, supporting the above interpretation of the origin of its polarization (see Fig.~\ref{fig:host} in Appendix~\ref{sec:app1}). The polarization angle in SN~2022ame, on the other hand, does not match any large-scale structure at the location of the SN. 
Even though the origin of the alignment/misalignment between the global galaxy structure and the local magnetic field is not fully clear \citep[see, \textit{e.g.},][]{Beck2015, Beck2020}, this may be the result of, \textit{e.g.}, of a local perturbation of the magnetic field.

\begin{figure}[ht!]
\plotone{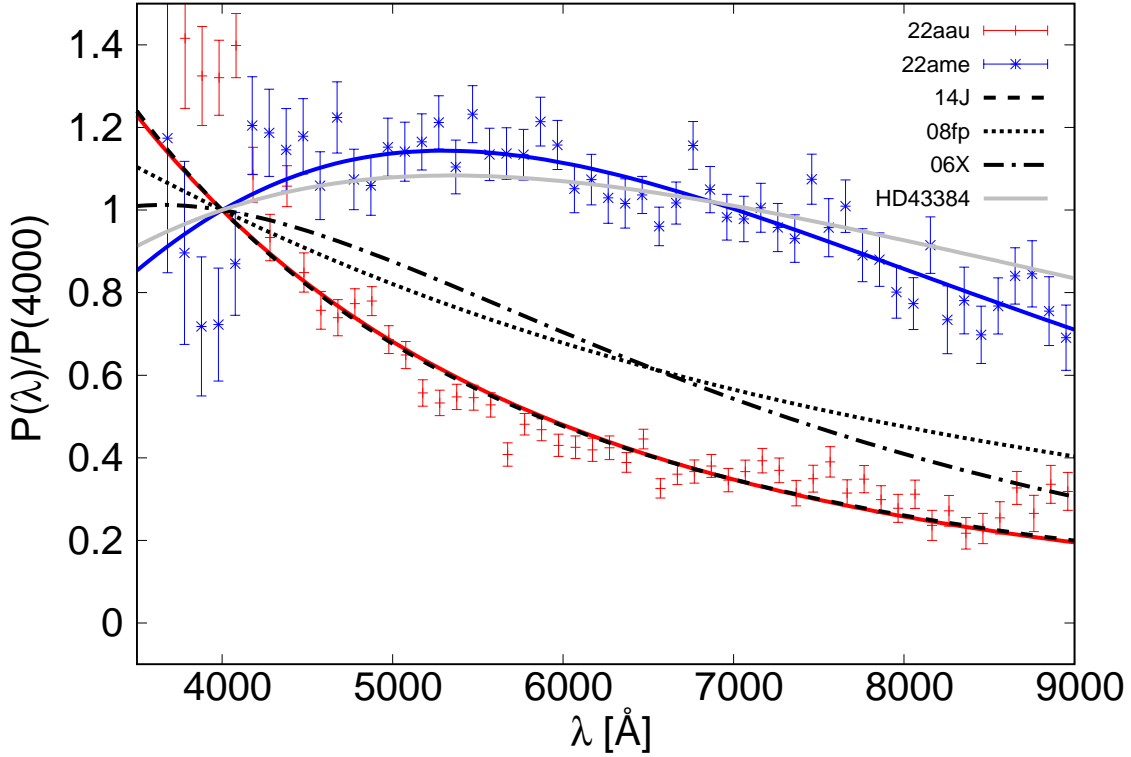}
\caption{Wavelength dependence of the polarization normalized at 4000 {\AA} toward SNe~2022aau (red) and 2022ame (blue) with their best-fit Serkowski curves. For comparison, the data of three Type~Ia SNe \citep[SNe~2014J, 2008fp and 2006X;][]{Patat2015} and a Galactic star \citep[HD~43384;][]{Cikota2018} are also plotted.}
\label{fig:wavelength_dep}
\end{figure}

Figure~\ref{fig:wavelength_dep} shows the wavelength dependence of the ISP toward SNe~2022aau and 2022ame, as compared with selected Type~Ia SNe and a Galactic star. The wavelength dependence of the ISP toward SN~2022aau deviates similarly from that of the MW as those toward some Type~Ia SNe, i.e., the polarization peaks are at a shorter wavelength ($\lambda_{\rm{max}}\lesssim4000$ {\AA}) than the typical ISP in the MW ($\lambda_{\rm{max}}\sim5500$ {\AA}). This is evidence for the presence of non-MW-like dust in its host galaxy, implying a significantly enhanced abundance of small grains compared to MW dust, as suggested for dust in the host galaxies of reddened Type~Ia SNe \citep[e.g.,][]{Patat2015, Chu2022}. This finding implies that such non-MW-like dust might be more common than expected in certain regions of galaxies, which might affect the picture of the star/planet formation in galaxies. On the other hand, the ISP of SN~2022ame is consistent with MW-like ISP, implying the existence of MW-like dust in its host galaxy. The two examples presented in this paper thus confirm that dust properties in external galaxies are diverse.

\begin{figure}[ht!]
\plotone{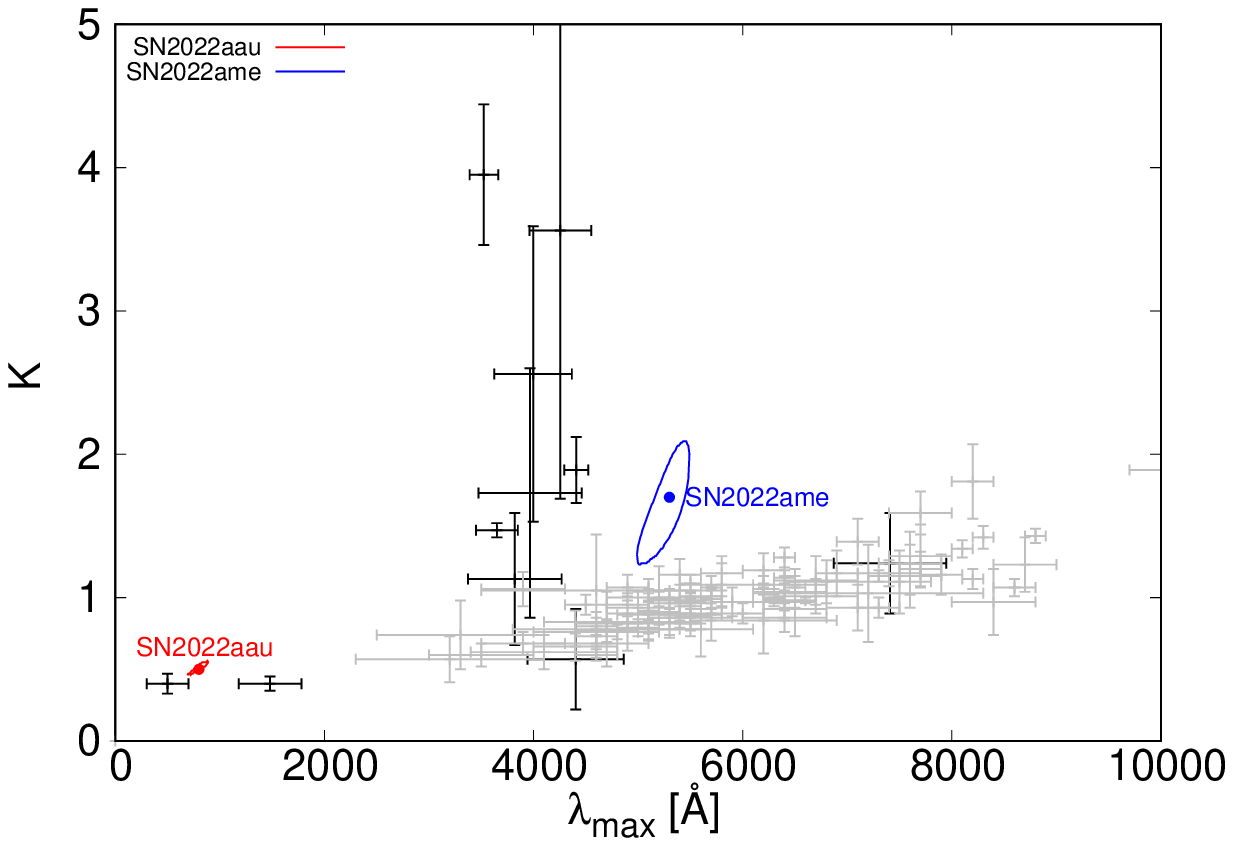}
\caption{The ISP $\lambda_{\rm{max}}$-$K$ diagram showing the Type~II SNe~2022aau (red) and 2022ame (blue) from this study. Several Type~Ia SNe \citep[black crosses;][]{Patat2015,Zelaya2017,Cikota2018} and a large number of MW stars \citep[gray crosses;][]{Whittet1992} are also included. The colored points show the best-fit values of the Type~II SNe, and the lines represent the 1-sigma confidence intervals for the fitting.}
\label{fig:k_l}
\end{figure}

We have derived the wavelength dependence of the ISP by fitting the polarization spectra with the Serkowski curve \citep[][]{Serkowski1975}:
\begin{equation}
    P(\lambda) = P_{\rm{max}} \exp \left[ -K \ln^{2} \left( \frac{\lambda_{\rm{max}}}{\lambda} \right) \right].
\end{equation}
The derived best-fit values for the parameters are $P_{\rm{max}}=13.72^{+4.23}_{-0.14}$ \%, $\lambda_{\rm{max}}=800^{+90}_{-110}$ {\AA} and $K=0.5^{+0.1}_{-0.1}$ for SN~2022aau and $P_{\rm{max}}=1.44^{+0.01}_{-0.06}$ \%, $\lambda_{\rm{max}}=5300^{+190}_{-310}$ {\AA} and $K=1.7^{+0.4}_{-0.4}$ for SN~2022ame. Here, from the empirical relation of \citet{Serkowski1975}, these values of $\lambda_{\rm{max}}$ imply $R_{V}\sim0.4$ and $\sim 2.9$ for SNe~2022aau and 2022ame, respectively. In Figure~\ref{fig:k_l}, the best-fit values and the 1-sigma confidence levels for the fitting are shown on the $\lambda_{\rm{max}}-K$ plane. SN~2022aau is located far from the cloud of the MW stars and close to some of Type~Ia SNe. SN~2022ame is close to the cloud of the MW stars, even though it still shows a slightly larger value of $K$ compared to the MW stars at more than 1 sigma confidence. It is noted that, since the best-fit value of $\lambda_{\rm{max}}$ for SN~2022aau is outside the wavelength range of our observations (3800-9200 {\AA}), the estimated values for SN~2022aau are not as reliable as those for SN~2022ame and the MW stars. However this does not affect the qualitative conclusion of a fundamental difference between SN~2022aau and the MW stars because the ISP peaks of the MW stars are caught by the observations.

Our results suggest that further investigation of IS dust properties using polarimetry of reddened SNe, in order to clarify the universality of such non-MW-like dust in other external galaxies, is highly promising. Furthermore, in order to identify the origin of such non-MW-like dust, it is important to study the dependence of the dust properties on environmental conditions such as gas density, strength of IS radiation field, strength of magnetic field, metallicity, etc.

\begin{acknowledgments}

This paper is based on observations collected at the European Southern Observatory under ESO programme 108.228K.001.
We are grateful to ESO's Paranal staff for the execution of the observations in service mode and the support astronomer, Paola Popesso, for the help with the arrangement of the observations.
The transient alert system developed by Steven Williams was helpful for our target selections.
T.N. thanks Masaomi Tanaka, Jian Jiang, Santiago Gonz{\'a}lez-Gait{\'a}n, Claudia P. Guti{\'e}rrez and Antonia Morales-Garoffolo for useful discussions.
T.N. is funded by the Academy of Finland project 328898.
T.N. acknowledges the financial support by the mobility program of the Finnish Centre for Astronomy with ESO (FINCA).
K.M. acknowledges support from the Japan Society for the Promotion of Science (JSPS) KAKENHI grant JP18H05223, JP20H00174,  and JP20H04737. The work is partly supported by the JSPS Open Partnership Bilateral Joint Research Project between Japan and Finland (JPJSBP120229923). 
H.K. is funded by the Academy of Finland projects 324504 and 328898.
M.B. acknowledges support from the Swedish Research Council (Reg. no. 2020-03330).

\end{acknowledgments}

%

\vspace{5mm}
\facilities{VLT (ESO)}





\appendix
\restartappendixnumbering

\section{The ISP angle in the host galaxies} \label{sec:app1}
\begin{figure}[ht!]
\epsscale{0.5}
\plotone{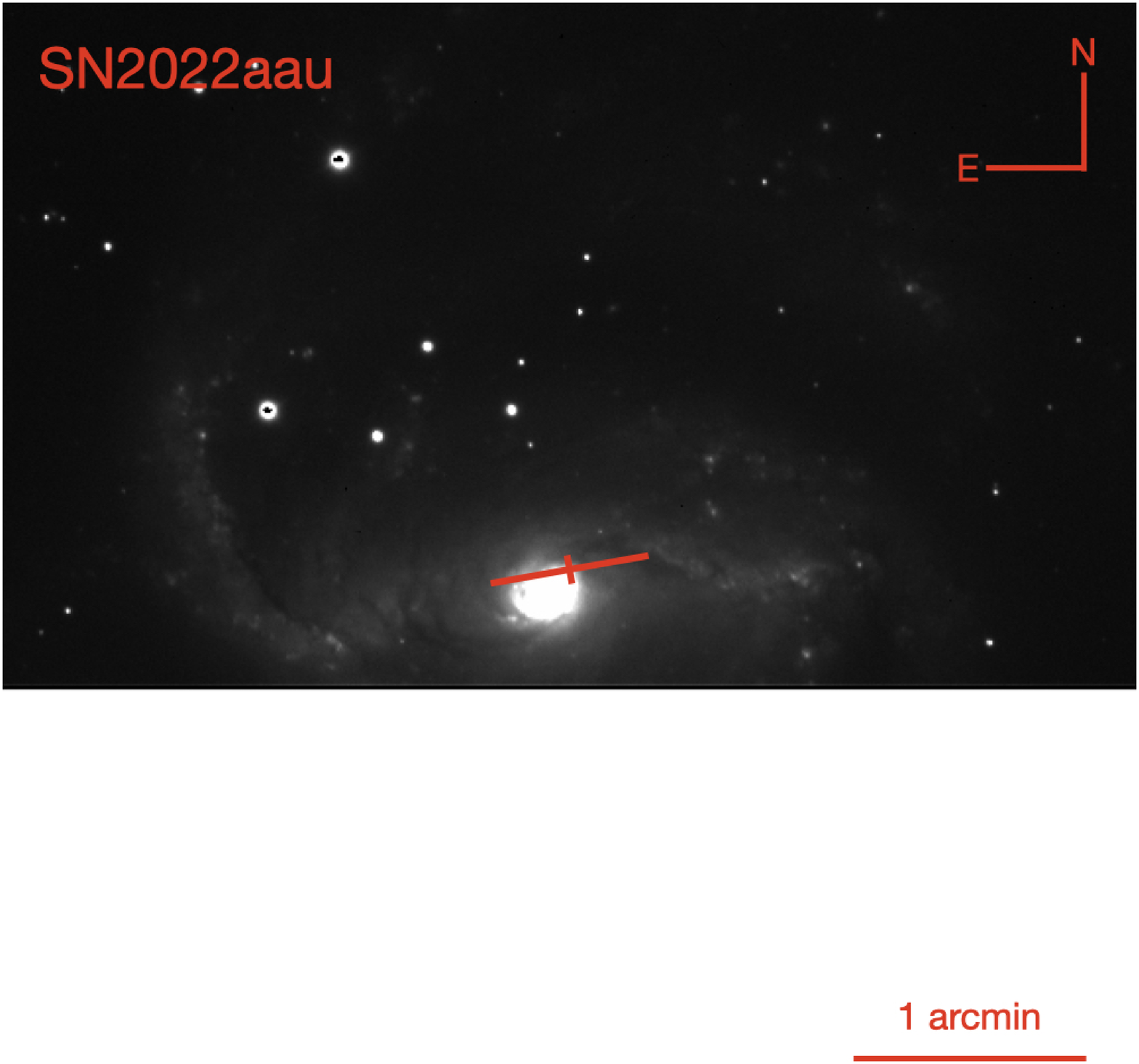}
\plotone{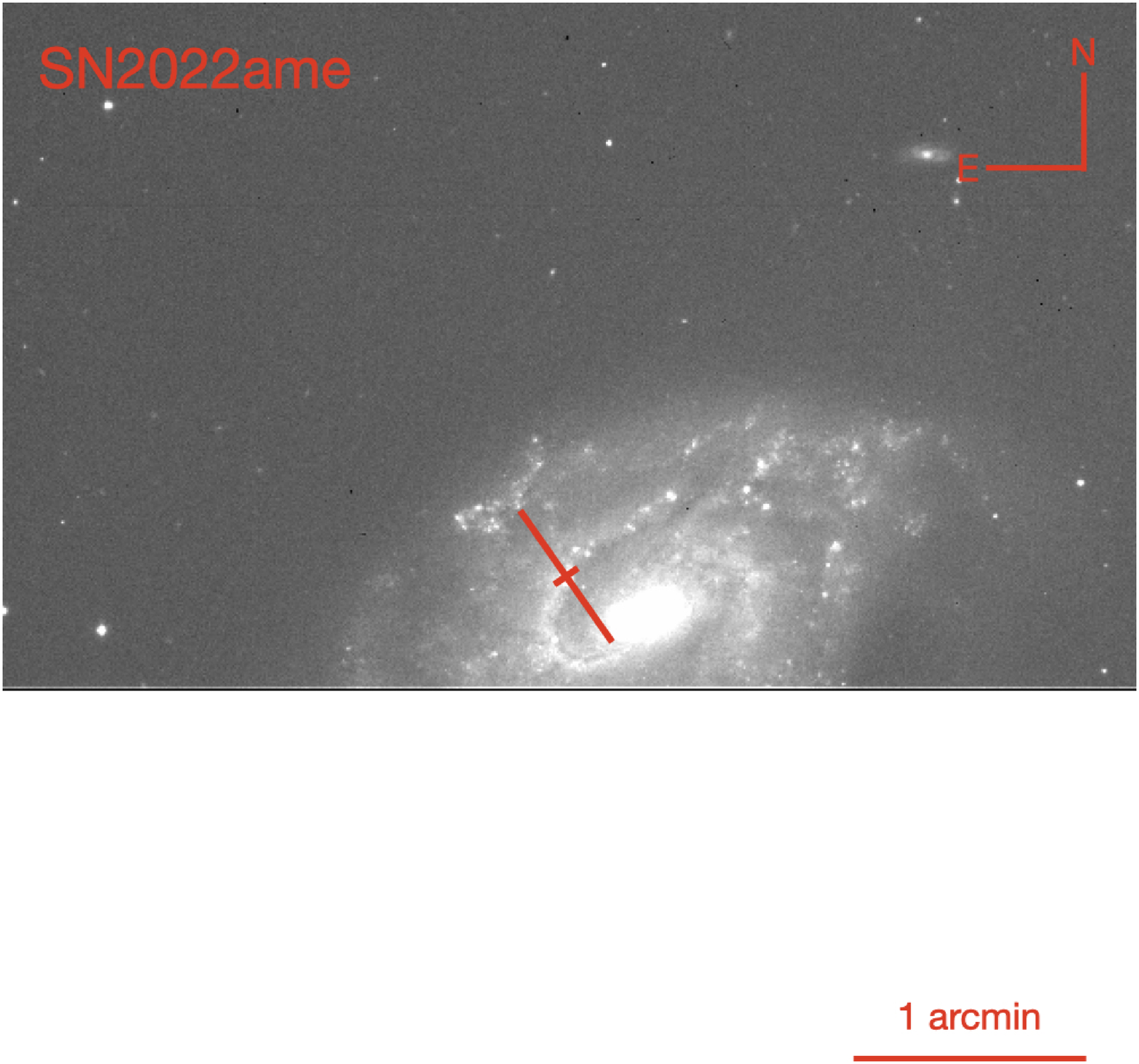}
\caption{VLT/FORS2 aquisition images of the SNe in this study and their host galaxies. The red crosses show the SN locations and the longer bars show the directions of the polarization angles of their ISP.}
\label{fig:host}
\end{figure}

The VLT/FORS2 aquisition images of the SNe (2022aau and 2022ame) and their host galaxies (NGC~1672 and NGC~1255, respectively) are shown in Fig.~\ref{fig:host}.

\restartappendixnumbering
\section{Spectra} \label{sec:app2}
\begin{figure}[ht!]
\plotone{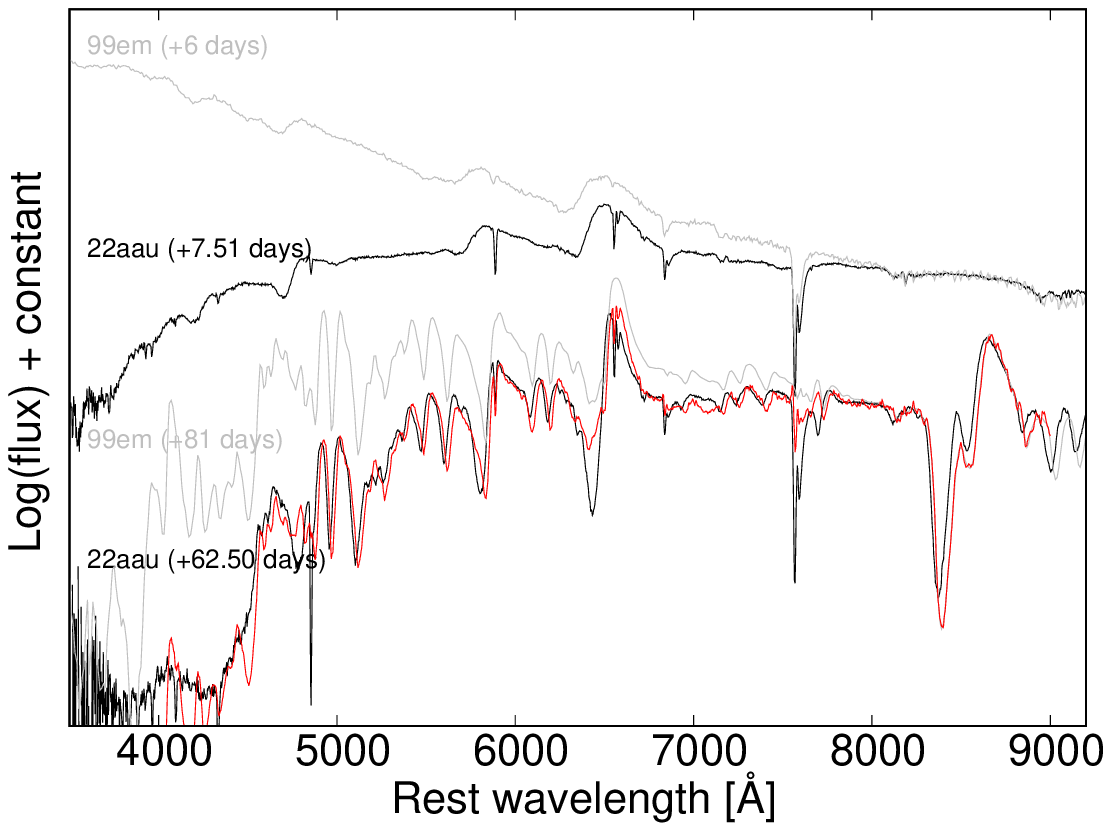}
\plotone{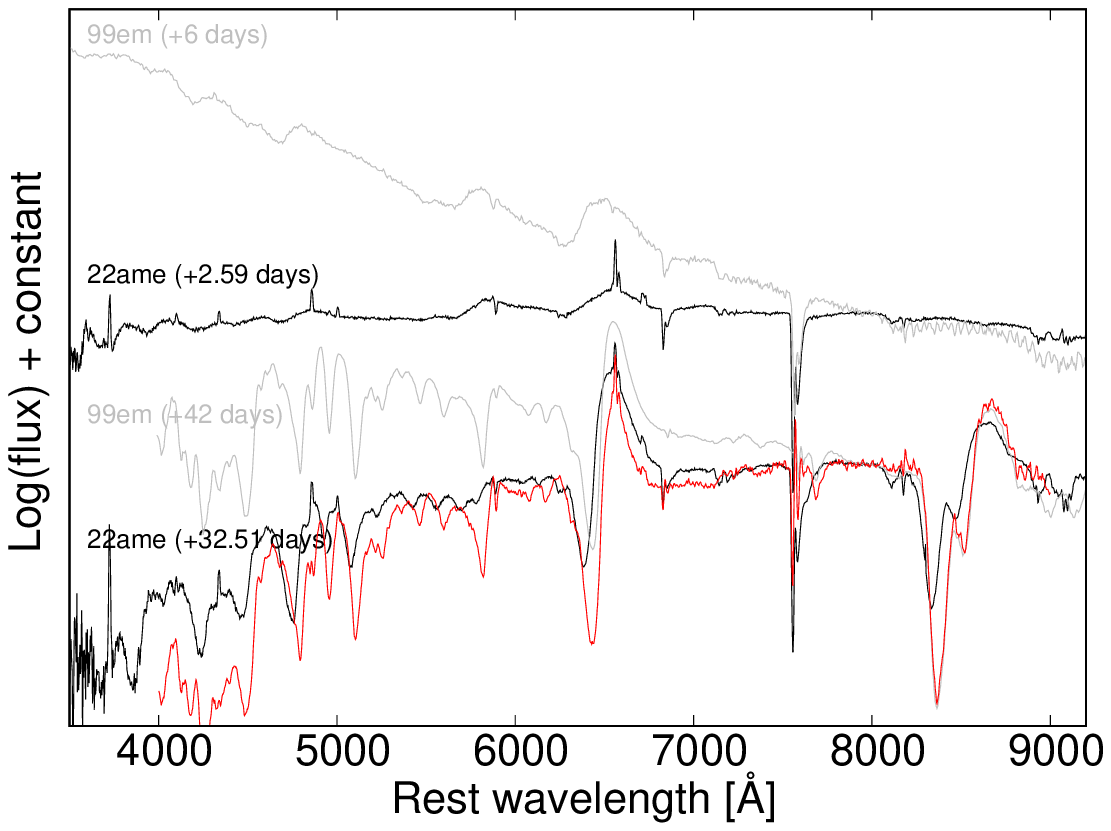}
\caption{The observed spectra of SNe~2022aau and 2022ame (black lines; without dust extinction correction), compared with those of the Type~IIP SN~1999em (gray lines; without dust extinction correction) taken from 
\citet[][]{Hamuy2001} and \citet[][]{Leonard2002}. The phases of SN~1999em are counted from the explosion date (24.1 October 1999) estimated by \citet[][]{Gutierrez2017}. The spectra of SN~1999em that are corrected by the estimated reddening from the first epochs toward SNe~2022aau and 2022ame, respectively, are shown with the red lines.}
\label{fig:spec}
\end{figure}

The VLT/FORS2 flux spectra of the SNe (2022aau and 2022ame) are shown in Fig.~\ref{fig:spec}. They are very reddened compared with those of SN~1999em, whose estimated dust extinction is $E(B-V)\sim0.1$ mag \citep[][]{Baron2000}. This also supports that SNe~2022aau and 2022ame suffer from substantial dust extinction.

Here, we investigate the time evolution of the reddening. We assume that the observed spectra of 99em at similar epochs (ignoring its extinction, which is small, i.e., E(B-V)$\sim 0.1$) is the original spectra of SN~2022aau and 2022ame before the extinction. Here, we scale the observed spectra of SN~1999em so that they have similar flux values with those of SNe~2022aau and 2022ame at the continuum regions around 8000 {\AA} (see Figure~\ref{fig:spec}). If we derive the reddening by comparing the spectra of SNe~2022aau and 1999em at the earlier phases (Phases +7.51 and +6 days, respectively) and apply this derived reddening to the spectrum of SN~1999em at latter epoch (Phase +81 days), this reddening-corrected spectrum of SN~1999em look similar to the spectra of SN~2022aau at the latter epoch (Phase +62.50 days; see Figure~\ref{fig:spec}). We obtain the same result for SN~2022ame as well (see Figure~\ref{fig:spec}). This implies that the reddening does not change with time in both cases of SNe~2022aau and 2022ame.

\begin{figure}[ht!]
\plotone{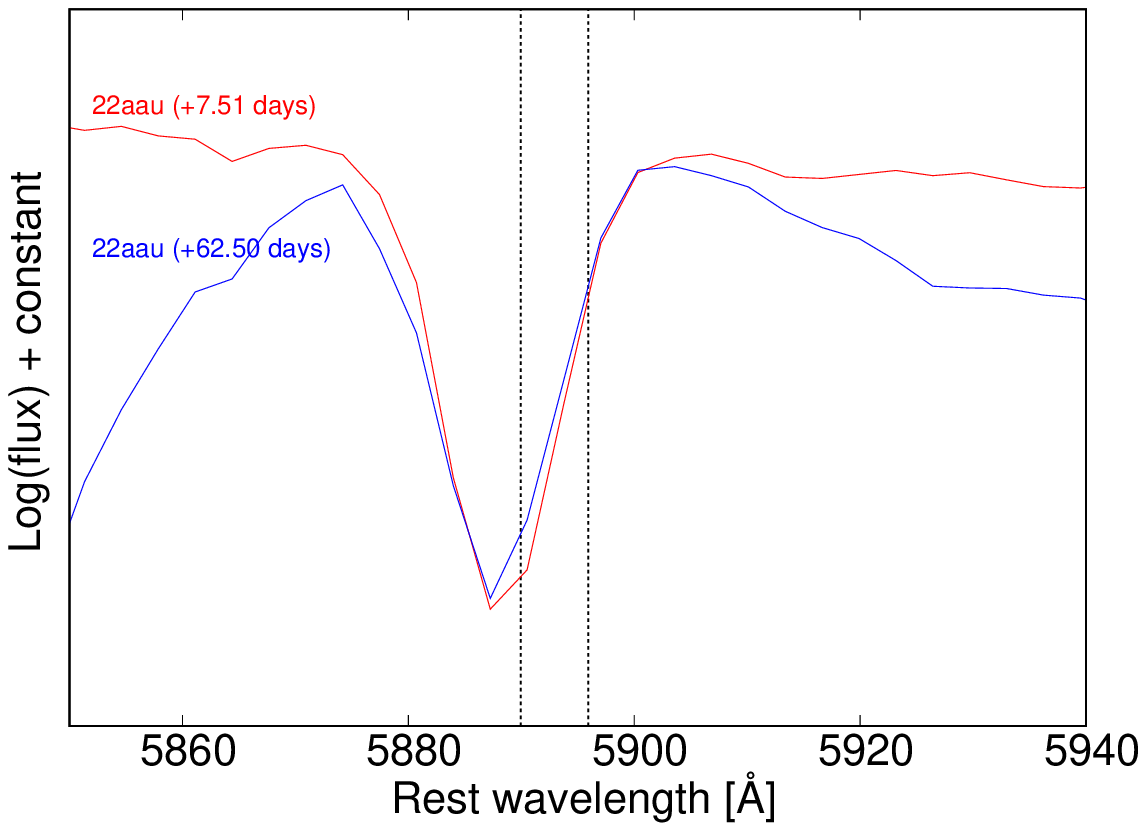}
\plotone{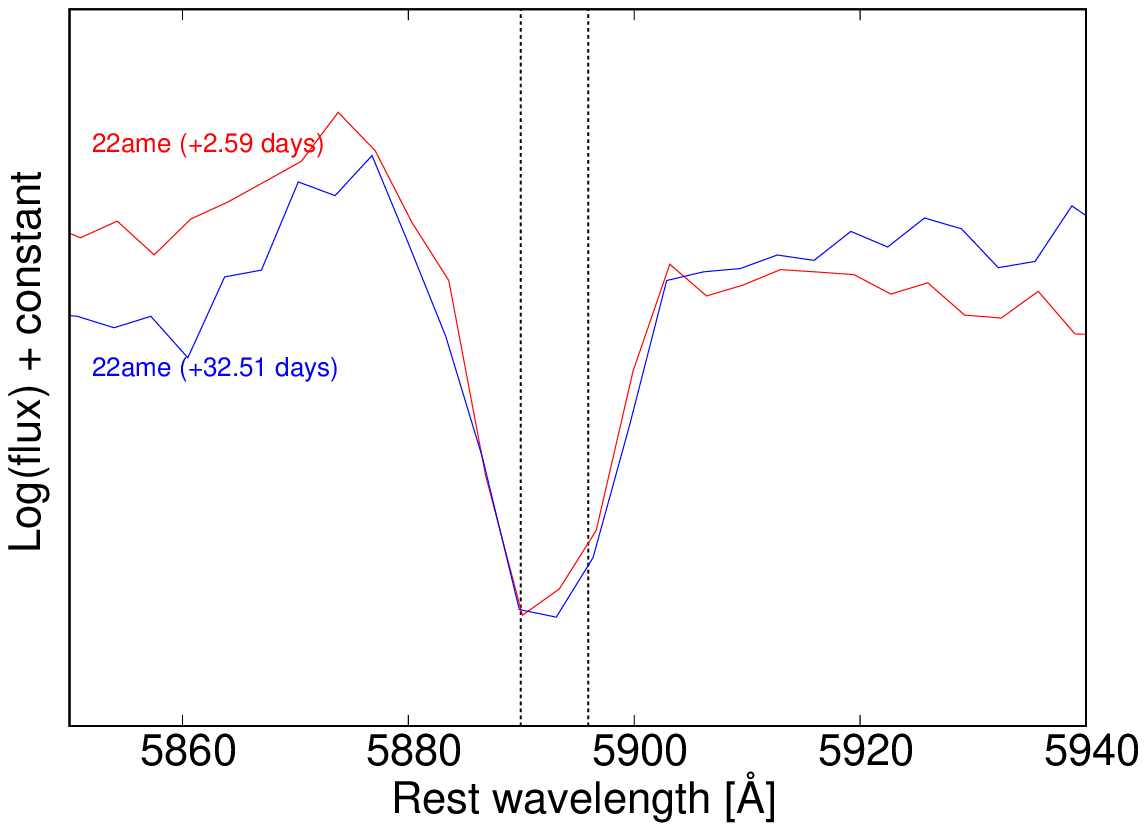}
\caption{
The Na~I~D absorption lines of SNe~2022aau and 2022ame at the redshift of their host galaxy. The vertical dotted lines indicate the wavelength of the Na~I~D lines (5889.950 and 5895.924 {\AA}).}
\label{fig:Na}
\end{figure}

Furthermore, the strength of the Na~I~D absorption lines, which should originate from the gas that contributes to the extinction, are time-independent in both cases (see Figure~\ref{fig:Na}).
There facts support that the dust that contributes to the ISP should be located in not a CS scale but a IS scale.

\bibliography{main}{}
\bibliographystyle{aasjournal}



\end{document}